# Modelling the Guaranteed QoS for Wireless Sensor Networks: A Network Calculus Approach

**Lianming Zhang[1], Jianping Yu[2], and Xiaoheng Deng[3]**

[1]*College of Physics and Information Science, Hunan Normal University, Changsha, Hunan, 410081, China*
[2]*College of Mathematics and Computer Science, Hunan Normal University, Changsha, Hunan, 410081, China*
[3]*Institute of Information Science and Engineering, Central South University, Changsha, Hunan, 410083, China*

Correspondence should be addressed to Lianming Zhang, lianmingzhang@gmail.com

**Abstract:** Wireless sensor networks (WSNs) became one of the high technology domains during the last ten years. Real-time applications for them make it necessary to provide the guaranteed Quality of Service (QoS). The main contributions of this paper are a system skeleton and a guaranteed QoS model that are suitable for the WSNs. To do it, we develop a sensor node model based on virtual buffer sharing and present a two-layer scheduling model using the network calculus. With the system skeleton, we develop a guaranteed QoS model, such as the upper bounds on buffer queue length/delay/effective bandwidth, and single-hop/ multi-hops delay/jitter/effective bandwidth. Numerical results show the system skeleton and the guaranteed QoS model are scalable for different types of flows, including the self-similar traffic flows, and the parameters of flow regulators and service curves of sensor nodes affect them. Our proposal leads to buffer dimensioning, guaranteed QoS support and control in the WSNs.

## 1. INTRODUCTION

Wireless sensor networks (WSNs) have been became one of the high technology domains of the seven seas, and theoretic and applications study about them are more and more regarded in recent years [1, 2, 3]. Real-time application areas for the WSNs encompass tracking, environment scouting, forecasting and medical care. Sink nodes of the WSNs respond in time on needs, so data channel between sink nodes and sensor nodes must offer a guaranteed Quality of Service (QoS). It includes deterministic sending rate, transmission without loss, end-to-end delay with the upper bound, and so on [1]. The guaranteed QoS plays an important role in data transmission for the WSNs. For example, the end-to-end delay with the upper bound is one of the guaranteed services, whether the upper bound on end-to-end can get a guarantee is a key to provide the guaranteed QoS and to complete effectively routing, congestion control and load balancing. To fulfill aims, the WSNs need to send some special probe packets [4]. The extra cost accounts for much total power under constrained energy, bandwidth and buffer size of a sensor node. However, it results in shortening of the WSNs' lifetime, and it is an important to provide the guaranteed QoS model and the performance evaluation method for the WSNs.

Network calculus is a set of recent developments that enable the effective derivation of deterministic performance bounds in networking [5, 6]. Compared with some traditional statistic theories, network calculus has the merit that provides deep insights into performance analysis of deterministic bounds. Now, research areas for the network calculus include mostly QoS control, resource allocation and scheduling, and buffer/delay dimensioning in the virtual circuit switched networks, the guaranteed service networks and the aggregate scheduling networks [5].

In recent years, the end-to-end delay bounds, in FIFO-multiplexing tandems, were estimated based on the least upper delay bound (LUDB) method [7]. The delay of individual traffic flows, in feed-forward networks under arbitrary multiplexing, was computed [8]. The maximum end-to-end delay, for a given flow in any feed-forward network under blind multiplexing, was calculated [9]. Resource allocation and congestion control was investigated in distributed sensor networks using the network calculus [10]. An analytical framework was presented, based on the network calculus, to analyze worst-case performance and to dimension the resource of sensor networks [11, 12, 13, 14]. The power management problem in video sensor networks was investigated [15]. The worst-case performance of the WSNs was analyzed [16]. Recently, the cluster-tree WSNs were modelled and dimensioned in the network calculus [17, 18, 19].

In previous papers [20, 21, 22, 23], we drawn the deterministic performance bound on end-to-end delay jitter for self-similar traffic regulated by a fractal leaky bucket regulator in a generalized processor sharing system, and got the deterministic and statistical performance bounds on end-to-end delay in the WSNs and the wireless mesh networks.

In this paper, we describe a generalized scenario of the WSNs, and present a practicable model of sensor nodes for the guaranteed service support using a scheme based on virtual buffer sharing. On the basis of the notion of flows and micro-flows, we propose, using arrival curves and service curves in the network calculus, a two-layer scheduling model for sensor nodes. We develop a guaranteed QoS model, including the upper bounds on buffer queue length/delay/effective bandwidth, and single-hop/multi-hops delay/jitter/effective bandwidth. Combined with the research results of predecessor researchers, the main different contributions of our work in this paper are as follows. Firstly, we present a system skeleton and a guaranteed QoS model that are suitable for the WSNs with some characteristics of the distribution and the multi-hops, and the sensor node model which not only fulfills these wants, but also makes performance analysis simpler. Secondly, we find that quantitative relations between the upper bounds on buffer queue length/delay/effective bandwidth, and single-hop/multi-hops delay/jitter/effective bandwidth and the service rate, the latency of the service curves in sensor nodes, and as well as the hops. Thirdly, we reveal the import of the service rate, the latency and the parameters of the regulators, including the Hurst parameter of self-similar traffic flows, on the guaranteed QoS. The findings' contributions are used to modelling the guaranteed QoS for the WSNs, and they may have potential applications to buffer and delay dimensioning, QoS support, routing implementing, congestion control and load balancing for the WSNs and other wireless networks with some characteristics of the distribution and the multi-hops.

The rest of the paper is organized as follows. Section 2 devotes to the background knowledge of the network calculus. Section 3 discusses a system skeleton, including a generalized scenario of the WSNs, a sensor node model, the flow source model, the guaranteed QoS service and the scheduling model of a sensor node. Section 4 draws the upper bounds on the guaranteed QoS model. Section 5 shows the numerical results and compares one another to demonstrate the availability and merits of the proposed skeleton, the guaranteed QoS model and our approach through same examples. Finally, Section 6 contains the summary of the results, some inferring remarks and future works.

## 2. BACKGROUND ON NETWORK CALCULUS

In this section, we provide a brief background on the network calculus used in the paper. Network calculus is the results of the studies on traffic flow problems, min-plus algebra and max-plus algebra applied to qualitative or quantitative analysis for networks in recent years, and it belongs to tropical algebra and topical algebra.

Network calculus can be classified into two types: deterministic network calculus and statistical network calculus. The former, using the arrival curve and the service curve, is mainly used to get the value of the exact solution of the bounds on network performance, such as queue length and queue delay. And then the latter, based on the arrival curve and the effective service curve, is used to get the stochastic or statistical bounds on network performance. Here we give only the necessary introductory material used in this paper.

*Theorem 1* (Queue Length and Queue Delay): Assume a flow passes through a sensor node, and the sensor node has an arrival curve $\alpha(t)$ and offers a service curve $\beta(t)$. The queue length $Q$ and queue delay $D$ of the flow, passing through the sensor node, satisfy respectively

$$Q \le \sup_{t\ge 0}\{\alpha(t)-\beta(t)\}, \qquad (1)$$

and

$$D \le \inf_{t\ge 0}\{d \ge 0 : \alpha(t) \le \beta(t+d)\}. \qquad (2)$$

*Theorem 2* (Multi-hops Service Curve): Assume a flow passes through the sensor node 1, node 2, …, node $N$ in sequence. Assume the sensor nodes offer service curves of $\beta^{(1)}$, $\beta^{(2)}$, …, $\beta^{(N)}$ to the flow, respectively. The fixed delays between two neighbor sensor nodes are $d_1$, $d_2$, …, $d_{N-1}$ in sequence. The multi-hops service curve $\beta^{m\text{-}h}$ satisfies

$$\beta^{m-h} = \beta^{(1)} \otimes \beta^{(2)} \otimes \ldots \otimes \beta^{(N)} \otimes \delta_{d_1+\ldots+d_{N-1}}, \qquad (3)$$

where $\otimes$ is the operator of the min-plus convolution given by

$$(f \otimes g)(t) = \begin{cases} \inf_{s\in[0,t]}[f(t-s)+g(s)], & t \ge 0 \\ 0, & t < 0 \end{cases},$$

and $\delta_d$ is called a burst delay function. For $0 \le t \le d$, $\delta_d(t)=0$, and for $t>d$, $\delta_d(t)=+\infty$.

In Eq. (3), we get, setting $n$=2, the single-hop service curve $\beta^{s\text{-}h}$ as shown

$$\beta^{s-h} = \beta^{(1)} \otimes \beta^{(2)} \otimes \delta_{d_1}.$$

The proof of the Theorems and more information about the network calculus is found in Refs. [5, 6].

## 3. SYSTEM SKELETON

### 3.1. System Model

In the following, we firstly describe a generalized scenario of the WSNs, where includes sink nodes, sensor nodes and a sensor field in Figure 1.

When certain sensor node of the sensor field probes an occurring event, the sensor node sends probed data to one of its neighbor sensor nodes according to the route arithmetic arranged in advance. And then the neighbor sensor node sends the data to one of its neighbor sensor nodes. Finally, the data probed by the first sensor node is transmitted to a sink node passing multi-hops.

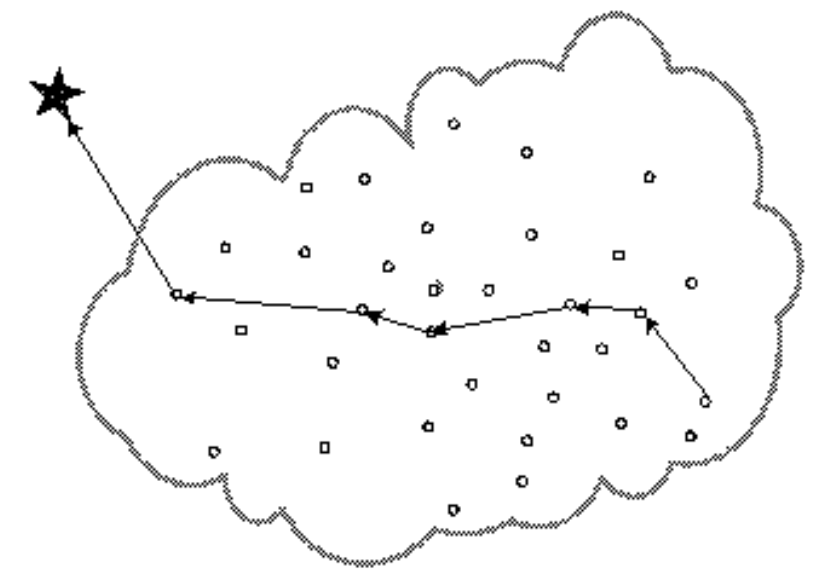

FIGURE 1: A generalized scenario of WSNs

In general, the energy of a sensor node is supplied by battery under constrained energy, so the storage and communication capacity of a sensor node is constrained. It is essential to provide the guaranteed QoS to lessen spending and to prolong a network lifetime.

The next, we present, using a scheme based on virtual buffer sharing, a sensor node model in Figure 2. The buffer of the sensor node is allocated to data channels between the sensor node and its upstream neighbor nodes. The probed data from its upstream neighbor nodes share the buffer of the sensor node. The scheduler of the sensor node sends the data to the downstream neighbor nodes according to the QoS priority. Figure 2 shows the case for the sensor node *j* and *i* upstream neighbor nodes, including the sensor node 1, node 2, …, node *i*.

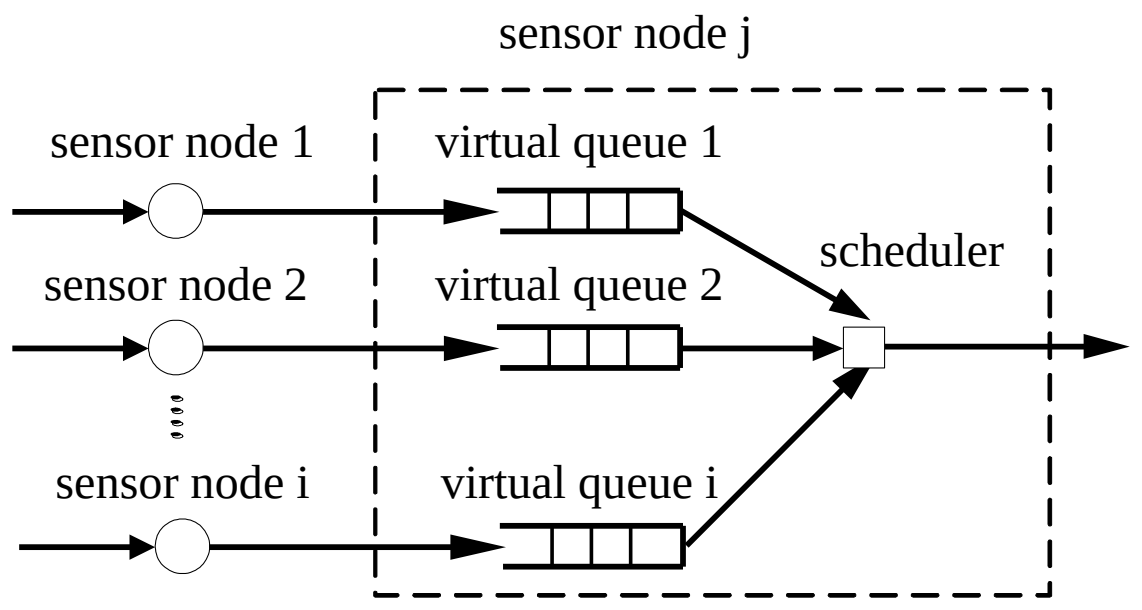

FIGURE 2: Sensor node model

*Remark 1*: The sensor node model using virtual buffer sharing has some merits as follows.

(1)The model provides a minimum guaranteed service rate for every data channel from upstream neighbor nodes under constrained bandwidth. Namely, when a data flow passes through a sensor node, the node guarantees a minimum service rate.

(2)The buffer and the bandwidth of the sensor node are shared by all of upstream neighbor nodes and delivered to them in part to their weights, so the WSNs get a larger gain from statistical multiplexing of the independent flows.

(3)The model makes performance analysis simpler, and it is suitable for mobile sensor nodes in the WSNs.

### 3.2. Flow Source Model

The dynamic and complexity property of the network and the fluctuation of the traffic possibly cause the burstiness of the traffic flows in the WSNs. They increase the average delay and result in the unfairness of resource allocation. It becomes more difficult in providing or analyzing the guaranteed QoS. In this paper, we can categorize traffic flows into two types: flow and micro-flow. The former contains file flows, audio flows and video flows, and so on. The latter, belonging to the identical type, aggregates a flow. The aggregate flow enters a sharing buffer to queue and schedule for the sensor node. In this paper, we select the leaky bucket source model due to its simplicity and practical applicability, and use leaky bucket regulators to regulate the micro-flows at every sensor node, to enable non-rule micro-flows to be restraint under the certain conditions. The micro-flow, regulated by the leaky bucket regulator, is indicated by envelope $\alpha(t)$ as shown in Eq. (4),

$$\alpha(t) = \min_{m\in\{1,\ldots,M\}} \{r^m \cdot t + b^m\}, \forall t \geq 0, \qquad (4)$$

where the case of $M$=1 agrees to the simple leaky bucket regulator, $b$ is interpreted as the burst parameter, and $r$ as the average arrival rate.

*Remark 2*: the micro-flow in an interval [$t$, $t$+$\tau$] is denoted by $A(t, t+\tau)$, and it has the following properties in Ref. [24].

*Property 1* (Additivity):

$$A(t_1, t_3) = A(t_1, t_2) + A(t_2, t_3), \forall t_3 > t_2 > t_1 > 0.$$

*Property 2* (Sub-additive Bounds):

$$A(t, t+\tau) \leq \alpha(\tau), \forall t \geq 0, \forall \tau \geq 0.$$

*Property 3* (Independence): All micro-flows are independent.

### 3.3. Guaranteed QoS

The guarantee QoS provides the QoS guarantees which involve the stability of performance, the usability and reliability of calculation resources, as well as the rationality of calculation price, and so on. In this paper, we mainly discuss how to provide guarantees for the QoS, including the upper bounds on buffer queue length/delay/effective bandwidth, and the upper bounds on single-hop/multi-hops delay/jitter/effective bandwidth. It is important to limit the values of buffer queue length/delay/jitter to a sustainable level below the upper bound. For example, once the value of tracking or environment scouting delay is beyond a certain value, such as the upper bound on end-to-end delay in the WSNs, the accuracy of tracking and the effectiveness of environment scouting have sharply declined. The following Table 1 reports an example of guaranteed service that comes from the experimental results for a real-time tracking environment and scouting application in the cluster-tree WSNs based on IEEE 802.15.4/ZigBee protocol in Ref. [19].

TABLE 1: An example of the guaranteed QoS

| micro-flows | buffer queue length | multi-hops delay |
|---|---|---|
| 1 | ≤5.38Kb | ≤7.15ms |
| 2 | ≤3.07Kb | ≤7.25ms |
| 3 | ≤4.07Kb | ≤9.07ms |

### 3.4. Two-Layer Scheduling Model

In the following, we present a two-layer scheduling model of a sensor node in Figure 3. The process of the model is as follows. Firstly, the micro-flows, entering a sensor node, with the same or similar QoS are regulated by a leaky bucket regulator, which is given in Eq. (4), and serves for the arrival curve $\alpha(t)$ of the next buffer. The functions $\alpha(t)$ and $A(t, t+\tau)$ satisfy Property 2. Secondly, we assume the FCFS (first come, first served) strategy is adopted in the buffer, and the micro-flows, belonging to the same type, enter a special buffer assigned by the sensor node. Finally, the aggregate flows are scheduled in a way of a service curve $\beta(t)$. The service curve is shown as follows.

From Properties 1 and 3, and Figure 3, the aggregate flows $A_j(t, t+\tau)$ and micro-flows $A_{j,k}(t, t+\tau)$, $k$=1, 2, …, $n$ satisfy

$$A_j(t,t+\tau) = \sum\nolimits_{k=1}^{n} A_{j,k}(t,t+\tau), \quad \forall t,\tau > 0\,. \quad (5)$$

From Ref. [25], the equivalent envelope curve $\alpha_j(t)$ of the aggregate flows and the envelope curve $\alpha_{j,k}(t)$ ($k$=1, 2, …, $n$) of the micro-flows satisfy

$$\alpha_j(t) = \sum\nolimits_{k=1}^{n} \alpha_{j,k}(t), \quad \forall t > 0\,. \quad (6)$$

The service curve $\beta_i(t)$ of the flow $i$ is defined as

$$\beta_i(t) = \beta(t) - \sum_{k=1,k\neq i}^{n} \alpha_k(t-\theta_k), \quad \forall t > \theta \geq 0\,, \quad (7)$$

where $\beta(t)$ is interpreted as the service curve of the sensor node, $\alpha_k$ as the arrival curve of the buffer $k$, and $n$ as the number of the buffers in the sensor node.

In order to simplify the calculation, without loss of generality, we assume the service curve $\beta(t)$ of the sensor node is a rate-latency function $\beta_{R,T}(t)$ given by

$$\beta(t) = \beta_{R,T}(t) = R\cdot(t-T), \quad \forall t > T > 0\,, \quad (8)$$

where $R$ is interpreted as the service rate, $T$ as the latency. Obviously for $R$>0 and $0\leq t\leq T$, $\beta_{R,T}(t)=0$.

From Property 3, Eqs. (5) and (6), the simple leaky bucket regulator is used, and the envelope curve of the regulator is

$$\alpha_i(t) = \varepsilon_i(t) = \sum\nolimits_{k=1}^{n} (b_{i,k} + r_{i,k}t)\,. \quad (9)$$

From Eqs. (4) and (8), if $\sum\nolimits_{i=1}^{n} r_i < R$, the parameter $\theta_i$ is optimized, and we have

$$\theta_i = T + \sum\nolimits_{k=1,k\neq i}^{n} b_k / R, \; i=1, 2, \ldots, n.$$

Substituting $\theta_i$ into Eq. (7), and combining Eq. (6) with Eq. (9), we get

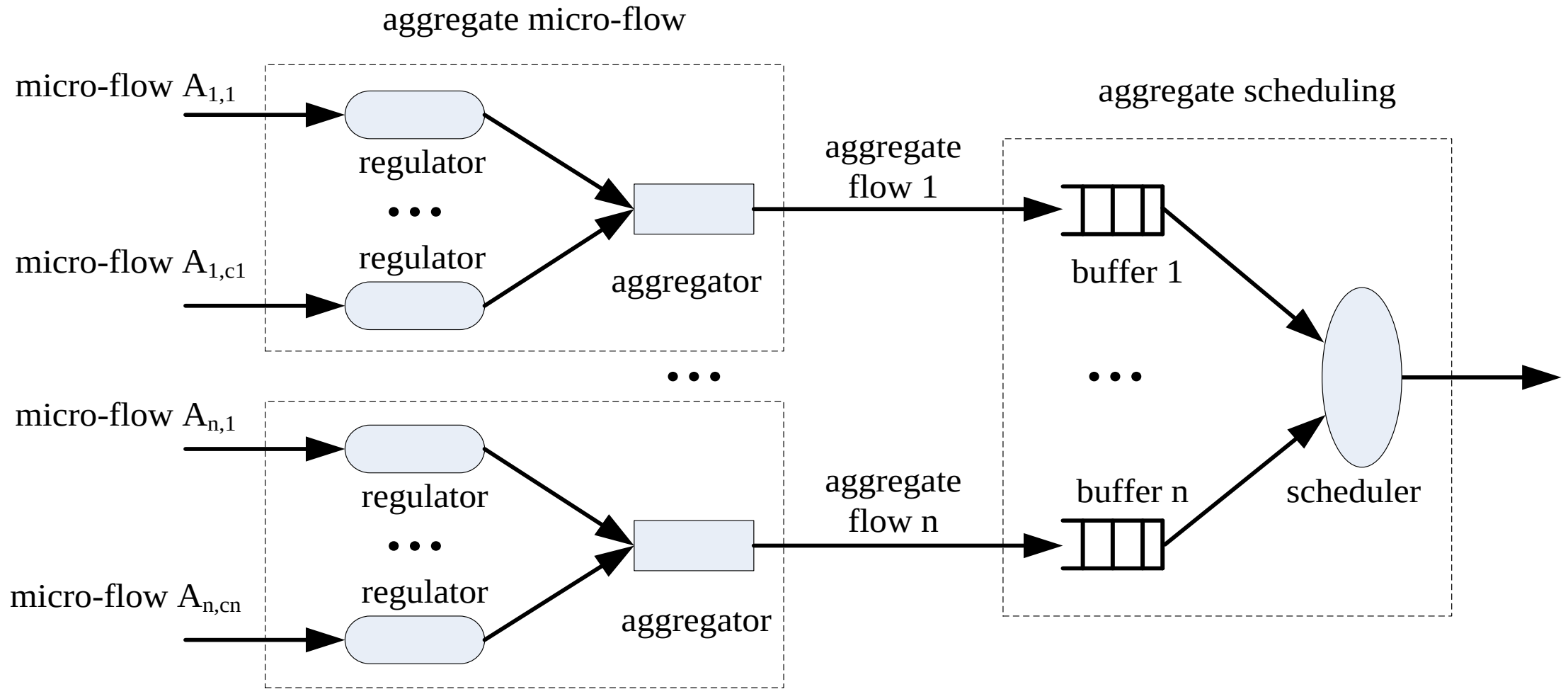

FIGURE 3: Two-layer scheduling model

$$\beta_i(t) = (R - \sum_{k=1,k\neq i}^{n} \sum_{j=1}^{c_j} r_{k,j}) \cdot (t - T - \sum_{k=1,k\neq i}^{n} \sum_{j=1}^{c_j} b_{k,j} / R) \quad (10)$$

*Remark 3*: From Eq. (10), we have known, each flow, which enters the scheduler, holds a certain service curve, and the service curve will not only be decided by the total service curve of the sensor node scheduler, but also by the arrival curve of the flow.

## 4. GUARANTEED QOS MODEL

In this section, we present, using the network calculus, the guaranteed QoS model. The model is mainly used in two aspects: one is the off-line dimensioning of a system, which is responsible for the quantification to get the pre-arranged resources providing the guarantee QoS; two is the on-line admission control, which is responsible to decide whether receives a new flow according to the QoS requirements and the usable resources. In the following, the guaranteed QoS model, including the upper bounds on $Q_i$, $D_i$, $e_i$, $DD_N$, $\Delta D_N$ and $ee_N$ (in Table 2) of the system skeleton in section 3 are discussed in the network calculus.

TABLE 2: The parameters of the QoS

| Symbol | Definition |
|---|---|
| $Q_i$ | buffer queue length of the sensor node *i* |
| $D_i$ | buffer queue delay of the sensor node *i* |
| $e_i$ | buffer queue effective bandwidth of the sensor node *i* |
| $DD_N$ | single-hop delay for $N$=2, and multi-hops delay for $N$>2 |
| $\Delta D_N$ | single-hop delay jitter for $N$=2, and multi-hops delay jitter for $N$>2 |
| $ee_N$ | single-hop effective bandwidth for $N$=2, and multi-hops effective bandwidth for $N$>2 |

### *4.1. Node QoS Model*

*Proposition 1:* (Upper Bound on Buffer Queue Length): In an interval [0, *t*], the upper bound on $Q_i$ satisfies

$$Q_i = \sup_{t\geq 0}\{\sum\nolimits_{k=1}^{n}(b_{i,k} + r_{i,k}t) - (R - \sum_{k=1,k\neq i}^{n} \sum_{j=1}^{c_j} r_{k,j}) \cdot (t - T - \sum_{k=1,k\neq i}^{n} \sum_{j=1}^{c_j} b_{k,j} / R)\} . \quad (11)$$

*Proof*: From Eq. (1), we have

$$Q_i \leq \sup_{t\geq 0}\{\alpha_i(t) - \beta_i(t)\} . \quad (12)$$

Substituting Eqs. (9) and (10) into Eq. (12), we hold

$$Q_i \leq \sup_{t\geq 0}\{\alpha_i(t) - \beta_i(t)\} = \sup_{t\geq 0}\{\sum\nolimits_{k=1}^{n}(b_{i,k} + r_{i,k}t) - (R - \sum_{k=1,k\neq i}^{n} \sum_{j=1}^{c_j} r_{k,j}) \cdot (t - T - \sum_{k=1,k\neq i}^{n} \sum_{j=1}^{c_j} b_{k,j} / R)\} .$$

*Proposition 2:* (Upper Bound on Buffer Queue Delay): In an interval [0, *t*], the upper bound on $D_i$ satisfies

$$D_i = T + \frac{\sum_{k=1}^{n} b_{i,k}}{R - \sum_{k=1,k\neq i}^{n} \sum_{j=1}^{c_j} r_{k,j}} + \sum_{k=1,k\neq i}^{n} \sum_{j=1}^{c_j} b_{k,j} / R . \quad (13)$$

*Proof*: From Eq. (2), we get

$$D_i \leq \inf_{t\geq 0}\{d \geq 0 : \alpha_i(t) \leq \beta_i(t+d)\} . \quad (14)$$

Substituting Eqs. (9) and (10) into Eq. (14), we have

$$D_i \leq \inf_{t\geq 0}\{d \geq 0 : \sum\nolimits_{k=1}^{n}(b_{i,k} + r_{i,k}t) \leq (R - \sum_{k=1,k\neq i}^{n} \sum_{j=1}^{c_j} r_{k,j}) \cdot (t + d - T - \sum_{k=1,k\neq i}^{n} \sum_{j=1}^{c_j} b_{k,j} / R)\} = \inf_{t\geq 0}\{d \geq 0 : d \geq T + \frac{\sum_{k=1}^{n}(b_{i,k} + r_{i,k}t)}{R - \sum_{k=1,k\neq i}^{n} \sum_{j=1}^{c_j} r_{k,j}} - t + \sum_{k=1,k\neq i}^{n} \sum_{j=1}^{c_j} b_{k,j} / R . \quad (15)$$

For $R \geq \sum_{k=1}^{n} \sum_{j=1}^{c_j} r_{k,j}$ , from Eq. (15), we get

$$D_i = T + \frac{\sum_{k=1}^{n} b_{i,k}}{R - \sum_{k=1,k\neq i}^{n} \sum_{j=1}^{c_j} r_{k,j}} + \sum_{k=1,k\neq i}^{n} \sum_{j=1}^{c_j} b_{k,j} / R .$$

*Proposition 3*: (Upper Bound on Buffer Effective Bandwidth): In an interval [0, *t*], the upper bound on $e_i$ satisfies

$$e_i = \sup_{t\geq 0} \frac{\sum_{k=1}^{n}(b_{i,k} + r_{i,k})}{t + D_i} , \quad (16)$$

where $D_i$ is given by Eq. (13).

*Proof*: Substituting Eq. (9) into Eq. (1.30) in Ref. [5], we have Eq. (16).

*Remark 4*: The leaky bucket regulators and aggregators don't increase the upper bounds on buffer queue length/ delay/effective bandwidth of a sensor node, and also don't increase the buffer requirements of the sensor node.

### 4.2. Single-Hop and Multi-hops QoS Model

*Proposition 4:* (Upper Bound on Single-Hop and Multi-hops Delay): Assume a flow passes through the sensor node 1, node 2, …, node $N$ in sequence, and the sensor node $i$ offers service curves of $\beta^{(1)}$, $\beta^{(2)}$, …, $\beta^{(N)}$ to the flow, respectively. The fixed delays between two neighbor sensor nodes are $d_1$, $d_2$, …, $d_{N-1}$ in sequence. The upper bound on $DD_N$ satisfies

$$DD_N = T_1 + \frac{\sum_{k=1}^{n} b_{i,k}^{(1)}}{\min\{R_1^{'},\ldots,R_N^{'}\}} + \sum_{i=1}^{N} T_i^{'} + \sum_{i=1}^{N-1} d_i \ , \qquad (17)$$

and

$$R_i^{'} = R_i - \sum_{k=1,k\neq i}^{n} \sum_{j=1}^{c_j} r_{k,j}^{(i)} \ ,$$

$$T_i^{'} = T_i + \sum_{k=1,k\neq i}^{n} \sum_{j=1}^{c_j} b_{k,j}^{(i)} / R_i \ .$$

Where $R_i$ and $T_i$ is interpreted as the service rate and the latency of the sensor node $i$, and $r_{k,j}^{(i)}$ and $b_{k,j}^{(i)}$ as the burst parameter and the average arrival rate of the leaky bucket regulator of the sensor node $i$, respectively.

*Proof*: From Eqs. (10), (3) and (8), we hold

$$\begin{aligned} \beta_N^{m-h} &= \beta_{\min\{R_1^{'},\ldots,R_N^{'}\},\sum_{i=1}^{N} T_i^{'}+\sum_{i=1}^{N-1} d_i} \\ &= \min\{R_1^{'},\ldots,R_N^{'}\} \cdot (t - \sum_{i=1}^{N} T_i^{'} - \sum_{i=1}^{N-1} d_i) \end{aligned} \quad .(18)$$

Substituting Eqs. (9) and (18) into Eq. (2), we have Eq. (17).

*Proposition 5* (Upper Bound on Single-Hop and Multi-hops Delay Jitter): Assume a flow passes through the sensor node 1, node 2, …, node $N$ in sequence, and the sensor node $i$ offers service curves of $\beta^{(1)}$, $\beta^{(2)}$, …, $\beta^{(N)}$ to the flow, respectively. The fixed delays between two neighbor sensor nodes are $d_1$, $d_2$, …, $d_{N-1}$ in sequence. The upper bound on $\Delta D_N$ satisfies

$$\Delta D_N = T_1 + \frac{\sum_{k=1}^{n} b_{i,k}^{(1)}}{\min\{R_1^{'},\ldots,R_N^{'}\}} + \sum_{i=1}^{N} T_i^{'} \ , \qquad (19)$$

where $T_1$ is interpreted as the latency of the first sensor node, $b_{i,k}^{(1)}$ as the burst parameter of the micro-flow $k$ of the flow $i$, entering the first sensor node, and others in Eq. (19) are shown in Eq. (17).

*Proof*: The upper bound on $DD_N$ got from Eq. (17) is the total delay, and the upper bound on $\Delta D_N$ and the fixed delay $D_c$ hold $\Delta D_N = DD_N - D_c$. The multi-hops fixed delay is defined as $D_c = \sum_{i=1}^{N-1} d_i$ . Therefore, Eq. (19) exists obviously.

*Proposition 6* (Upper Bound on Single-Hop and Multi-hops Effective Bandwidth): Assume a flow passes through the sensor node 1, node 2, …, node $N$ in sequence, and the sensor node $i$ offers service curves of $\beta^{(1)}$, $\beta^{(2)}$, …, $\beta^{(N)}$ to the flow, respectively. The fixed delays between two neighbor sensor nodes are $d_1$, $d_2$, …, $d_{N-1}$ in sequence. The upper bound on $ee_N$ satisfies

$$ee_N = \max\{r_{i.k}^{(1)}, b_{i.k}^{(1)} \Big/ (T_1 + \frac{\sum_{k=1}^{n} b_{i.k}^{(1)}}{\min\{R_1^{'},\ldots,R_N^{'}\}} + \sum_{i=1}^{N} T_i^{'} + \sum_{i=1}^{N-1} d_i)\} \ ,(20)$$

where the parameters in Eq. (20) are given in Eq. (19).

*Proof*: From Eqs. (9), (16), we get

$$ee_N \leq \max\{ r_{i,k}, b_{i,k} / D_{i,k} \} \ ,$$

for $D_{i,k} \geq D_i$, from Eqs. (17) and (16), we have Eq. (20).

*Remark 5*: The single-hop scenario is a special case of the multi-hops WSNs. In Eqs. (17), (19) and (20), we get the single-hop QoS model for $N$=2, and get the multi-hops QoS model for $N$>2.

*Remark 6*: The leaky bucket regulators and aggregators don't increase the upper bounds on single-hop/multi-hops delay/jitter/effective bandwidth of the WSNs.

## 5. NUMERICAL RESULTS

In this section, we give the numerical results to demonstrate the effectiveness and simplicity of our method. Without loss of generality, we research a general scenario of the WSNs as shown in Figure 4. If $N$=2, there is a single-hop case, otherwise, there is a multi-hops case. The two-layer scheduling model presented in Section 3 is used for all sensor nodes. The service curves $\beta(t)$ of the sensor nodes are given in Eq. (10), where $R$ is interpreted as the service rate and $T$ as the latency of the service curves of the sensor nodes. The fixed delay between two neighbor sensor nodes is marked by $d$.

Figure 4 shows the transmission of three flows in the WSN. The three flows: flow1, flow2 and flow3 are marked as $A_1(t)$, $A_2(t)$ and $A_3(t)$, respectively. They come from the sensor nodes A, B and C. Hence, without any loss of generality, we assume the flow $A_1(t)$ contains three micro-flows: $A_{1,1}(t)$, $A_{1,2}(t)$, $A_{1,3}(t)$, the flow $A_2(t)$ contains two micro-flows: $A_{2,1}(t)$ and $A_{2,2}(t)$, and the flow $A_3(t)$ contains one micro-flow: $A_{3,1}(t)$.

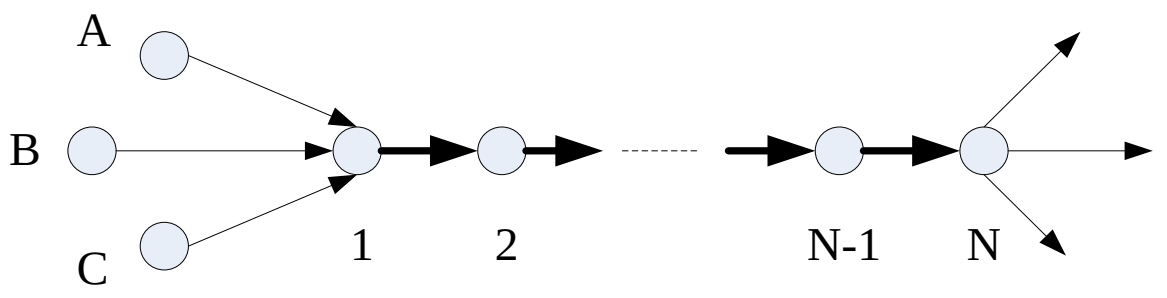

FIGURE 4: General scenario of WSNs

Recently research suggests that the sensory data flow is bounded by arrival curve $\alpha(t)$= 576(bps)+390(b)·$t$ in the cluster-tree WSNs based on IEEE 802.15.4/ZigBee protocol in Ref. [19]. Here we consider the case of $M$=2 in Eq. (4) and assume that every micro-flow is regulated by the leaky bucket regulator $\alpha(t)$ as shown in Eq. (9). The average arrival rate $r_{i,k}$ and the burst tolerance $b_{i,k}$ of the six micro-flows are shown in Table 3. Obviously, the arrival curves of the flows are given by Eq. (10).

TABLE 3: The parameters of the three flows

| flows $A_i(t)$ | mico-flows $A_{i,k}(t)$ | average arrival rate $r_{i,k}$(Kbps) | burst tolerance $b_{i,k}$(Kb) |
|---|---|---|---|
| $A_1(t)$ | 1 | 500 | 30 |
| | 2 | 300 | 300 |
| | 3 | 420 | 150 |
| $A_2(t)$ | 1 | 600 | 200 |
| | 2 | 240 | 500 |
| $A_3(t)$ | 1 | 300 | 200 |

*Remark 7*: The units of buffer queue length $Q$, effective bandwidth $e$ and $ee$ are Mb, the units of delay $D$ and $DD$, the time $t$, the latency $T$ and the fixed delay $d$ are ms and the unit of the service rate $R$ is Mbps except the units that are given.

### 5.1. Node QoS

In the following, we discuss the relations between the sensor node QoS and the parameters of the service curve provided by the sensor nodes, and the time evolution of the sensor node QoS.

Figure 5 shows the import of the service rate $R$ and the latency $T$ on the upper bounds on buffer queue length $Q$ and the evolution laws of $Q$ in a sensor node. We see a straightforward dependency: the upper bound on $Q$ is smaller for smaller service rate $R$ with low-value or smaller latency $T$; it is smaller for larger service rate $R$ with high-value or larger evolution time $t$. For all flows, the changing tendency of the upper bound on $Q$ with the increase of the service rate $R$ or the latency $T$ and the time evolution $t$ of $Q$ are the same. The size deviation of the upper bound on $Q_1$, $Q_2$ and $Q_3$ of the flows: $A_1(t)$, $A_2(t)$ and $A_3(t)$ is equal regardless of $R$ values and $T$ values. The upper bound on $Q_2$ of $A_2(t)$ is more than that of $Q_1$ of $A_1(t)$, and that of $Q_3$ of the $A_3(t)$ is smallest. Obviously, the impart of the latency $T$ or burst tolerance $b$ on the upper bound on $Q$ is more than that of the service rate $R$ or average arrival rate $r$, respectively.

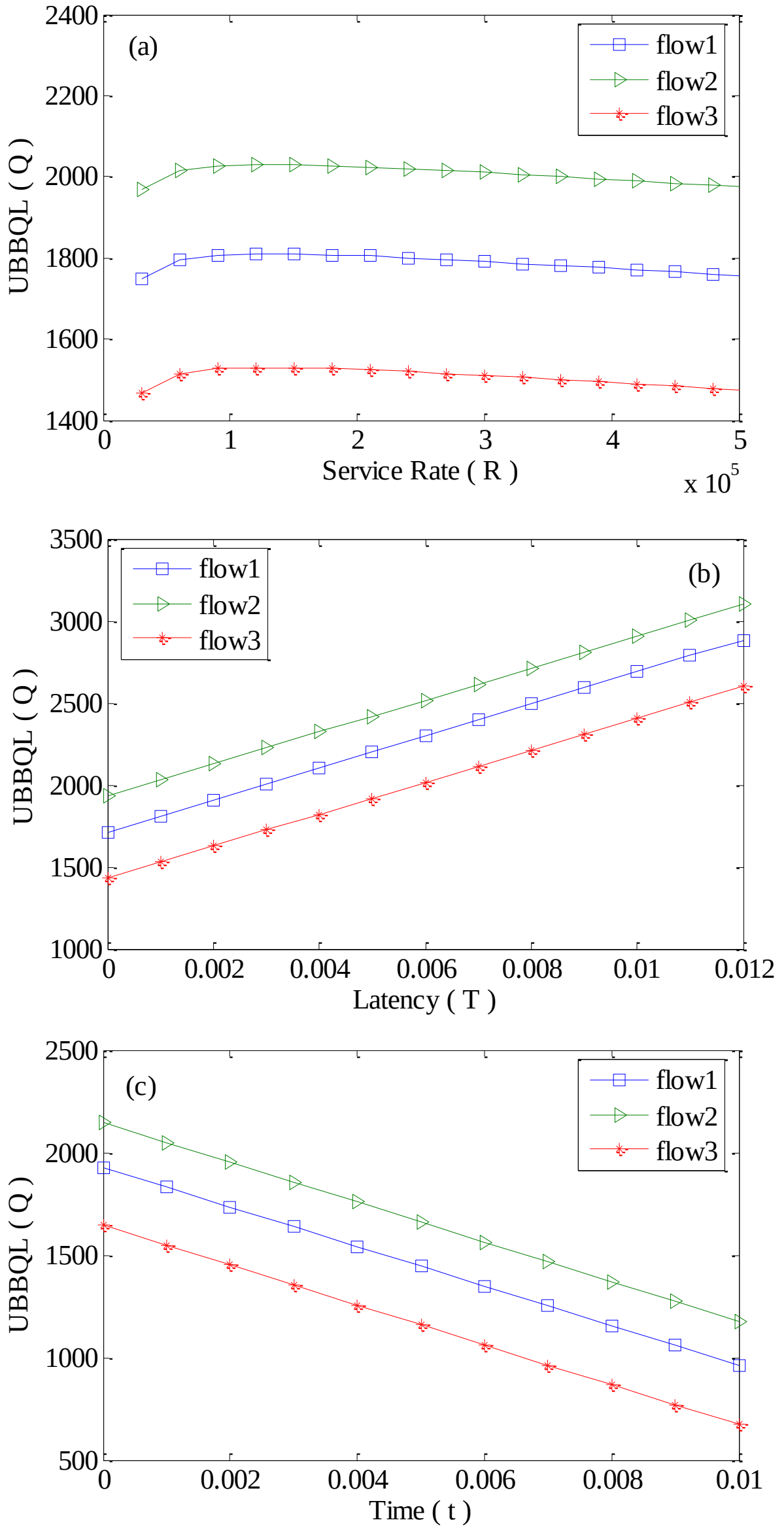

FIGURE 5: The upper bounds on buffer queue length (UBBQL) $Q$ in Kb of a sensor node: (a) $Q$ as a function of the service rate $R$ in Kbps for $T$=1 and $t$=1.2; (b) $Q$ as a function of the latency $T$ in s for $R$=100 and $t$=1.2; (c) $Q$ as a function of the evolution time $t$ in s for $R$=100 and $T$=1.

Figure 5(a) plots the $Q$ curves as a function of the service rate $R$. The upper bound on $Q$ for any flow reaches a maximum $Q_{max}$ when the service rate $R$=128, and The $Q_{max}$ value of the flows: $A_1(t)$, $A_2(t)$ and $A_3(t)$ is 1.81, 2.03 and 1.53, respectively. The shapes of curves at the two sides of the maximum point are asymmetric. For example, at the distance 50 from the maximum point on the left, the $Q_{max}$ value of the three flows is 1.81, 2.02 and 1.52 respectively. At the same distance on the right, the $Q_{max}$ value of the three flows is 1.81, 2.03 and 1.53, respectively.

Figure 5(b) plots the $Q$ curves as a function of the latency $T$. The upper bound on $Q$ increases linearly with the increase of the latency $T$, and the slope of each line is $9.764\times10^4$.

Figure 5(c) plots the $Q$ curves as a function of the evolution time $t$. There exists a linear relationship between the upper bound on $Q$ and the evolution time $t$ and the same slopes of the all lines are $-9.642\times10^4$.

Figure 6 shows the import of the service rate $R$ and the latency $T$ on the upper bounds on buffer queue delay $D$ in a sensor node. We see a straightforward dependency: the upper bound on $D$ is smaller for larger service rate $R$; it is smaller for smaller latency $T$.

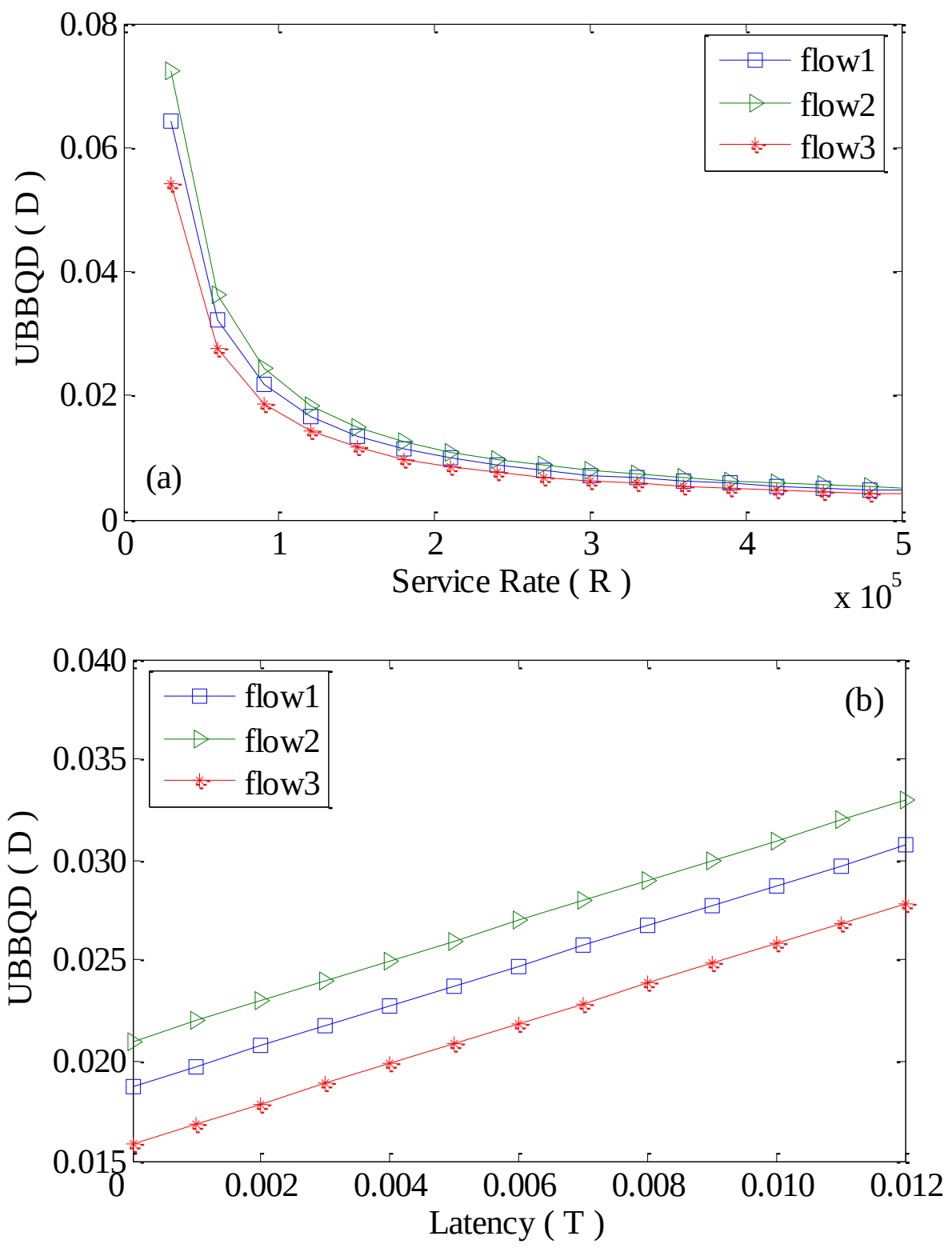

FIGURE 6: The upper bounds on buffer queue delay (UBBQD) $D$ in s of a sensor node: (a) $D$ as a function of the service rate $R$ in Kbps for $T$=1; (b) $D$ as a function of the latency $T$ in s for $R$=100.

Figure 6(a) plots the $D$ curves as a function of the service rate $R$. The $D$ values, curving inwards, decay with the increase of $R$ regardless of $T$ values, nearly converging 0 for all flows. The decay rates in the upper bounds on $D$ by the near exponential increase with the increase of the service rate $R$ for certain flow, and increase with the increase of the burst tolerance $b$ of the flows with the same service rate $T$. For instance, if $T$=1 and $R$=50, the $D$ value of the flows: $A_1(t)$, $A_2(t)$ and $A_3(t)$ is 38.7, 43.3 and 32.8, and if $T$=1 and $R$=200, the $D$ value of the three flows is 10.3, 11.4 and 8.9, respectively.

Figure 6(b) plots the $D$ curves as a function of the latency $T$. The upper bounds on $D$ increase linearly with the increase of the latency $T$ regardless of the $R$ values. The slopes of all lines are 1.

Figure 7 shows the import of the service rate $R$ and the latency $T$ on the upper bound on buffer effective bandwidth $e$ in a sensor node. We see a straightforward dependency: the upper bound on $e$ is larger for larger service rate $R$; it is larger for smaller latency $T$.

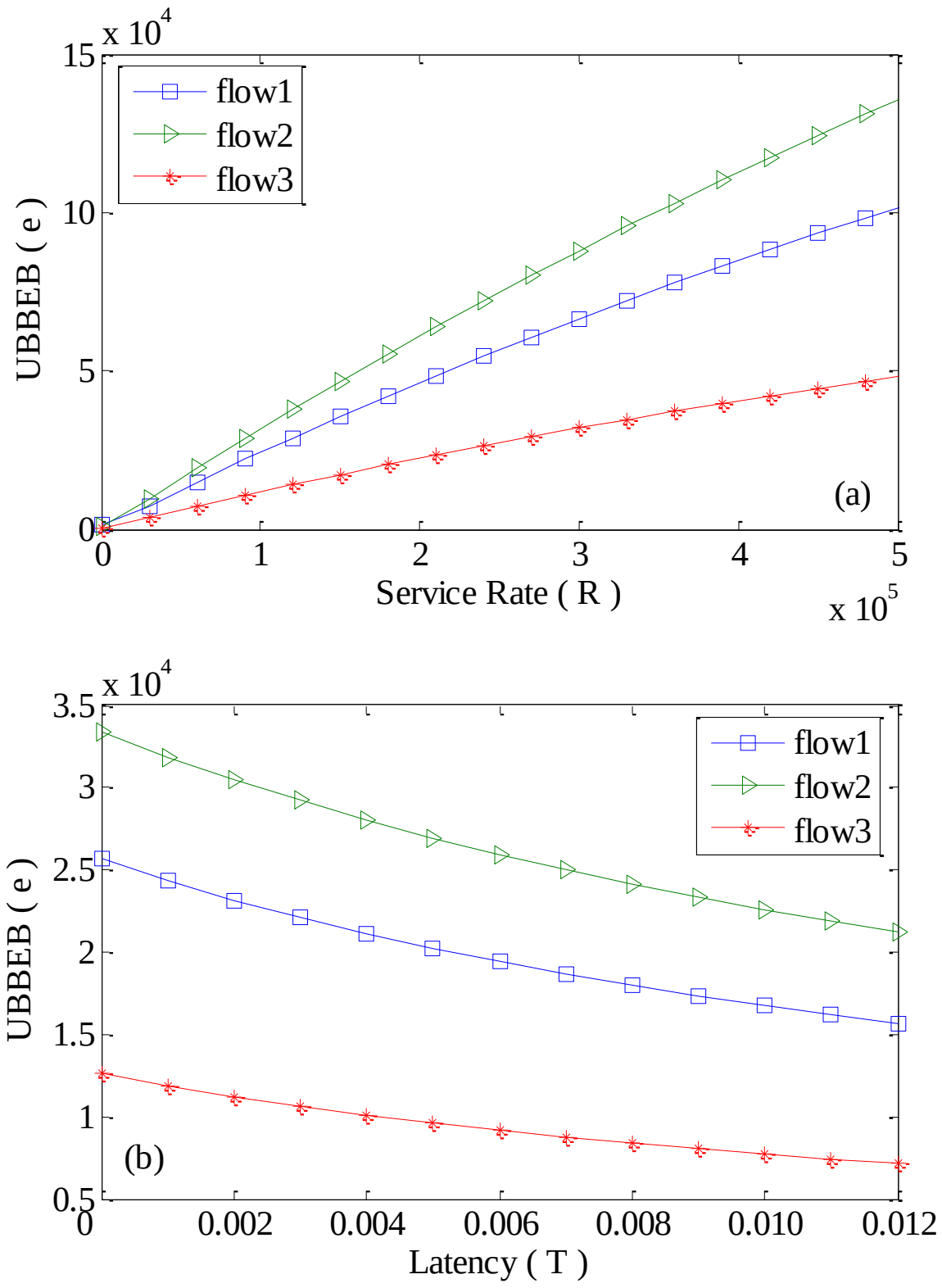

FIGURE 7: The upper bounds on buffer effective bandwidth (UBBEB) $e$ in Kb of a sensor node: (a) $e$ as a function of the service rate $R$ in Kbps for $T$=1; (b) $e$ as a function of the latency $T$ in s for $R$=100.

Figure 7(a) plots an $e$ curves as a function of the service rate $R$. The $e$ values increase with the increase of $R$ values, and the increase rate is getting smaller and smaller with the increase of $R$ values for certain flow regardless of the values of the latency $T$. The delay rates of the increase rates decrease with the increase of the burst tolerance $b$ of the flows. For instance, if $T$=1 and $R$=50, the $e$ value of the flows: $A_1(t)$, $A_2(t)$ and $A_3(t)$ is 12.41, 16.17 and 6.10, and if $T$=1 and $R$=200, the $e$ value of the three flows is 46.47, 61.18 and 22.44, respectively.

Figure 7(b) plots an $e$ curves as a function of the latency $T$. The upper bounds on $e$ decrease with the increase of $T$ values, and the decay rate is getting smaller and smaller with the increase of $T$ values for certain flow regardless of the $R$ values. The $e$ curves of all flows are near parallel.

In summary, the performance curves denote the upper bounds of the sensor node QoS. In Figures 5, 6 and 7, the curves show the deterministic worst-case length/delay/ effective bandwidth in the buffer queue of a sensor node, respectively. It means that the values of the buffer queue length/delay must are lower than the values of the performance curves. We can reduce, regulating the average arrival rate $r$ and the burst tolerance $b$ of the micro-flows by controlling the parameters of the regulators or regulating the service rate $R$ or the latency $T$ of a sensor node by controlling the parameters of the scheduler, the values of the upper bounds on buffer queue length/delay of a sensor node to achieve these purposes that the buffer queue length/delay is very small. Instead, we can increase, regulating the average arrival rate $r$ and the burst tolerance $b$ or regulating the service rate $R$ or the latency $T$, the value of the upper bound on buffer queue effective bandwidth to gain a guaranteed bandwidth for those flows through the sensor node, and eventually reduce the buffer queue delay.

### *5.2. Multi-hops and Single-Hop QoS*

In the following, we discuss the relations between the multi-hops QoS and the single-hop QoS and the parameters of the service curve provided by the sensor nodes and the hops. We still use the general scenario of WSNs as known in Figure 4.

#### *5.2.1. The case 1*

The sensor nodes (node 1, node 2, …, node $N$−1, node $N$) have the same service curve: $\beta_1(t)=\beta_2(t)=\ldots=\beta_{N-1}(t)=\beta_N(t)=\beta(t)=R(t-T)$. From Eq. (8), we have $R_1=R_2=\ldots=R_{N-1}=R_N=R$, and $T_1=T_2=\ldots=T_{N-1}=T_N=T$. To make easy the following discussion, we assume the fixed delays between two neighbor sensor nodes are the same: $d_1=d_2=\ldots=d_{N-1}=d$. Firstly, we investigate the multi-hops scenario with hops higher than 2.

Figure 8 shows the import of the service rate $R$, the latency $T$ and the hops $N$ on the upper bounds on multi-hops delay $DD$. We see a straightforward dependency: the upper bound on $DD$ is smaller for larger service rate $R$; it is smaller for smaller latency $T$ and smaller hops $N$.

Figure 8(a) plots a $DD$ curves as a function of the service rate $R$. The $DD$ values, curving inwards, decay with the increase of $R$ regardless of $T$ values, $N$ values and $d$ values for certain flow. The decay rates in the upper bounds on $DD$ by the near exponential increase with the increase of the service rate $R$ for certain flow, and increase slightly with the increase of the burst tolerance $b$ of the flows for the same $T$. For instance, if $T_1=T_2=\ldots=T_{N-1}=T_N=T=1$, $R_1=R_2=\ldots=R_{N-1}=R_N=R=50$, $d_1=d_2=\ldots=d_{N-1}=d=2$ and $N$=10, the $DD$ value of the flows: $A_1(t)$, $A_2(t)$ and $A_3(t)$ is 315, 320 and 309, and if $T$=1, $R$=200, $d$=2, and $N$=10, the $DD$ value of the three flows is 100, 102 and 99, respectively.

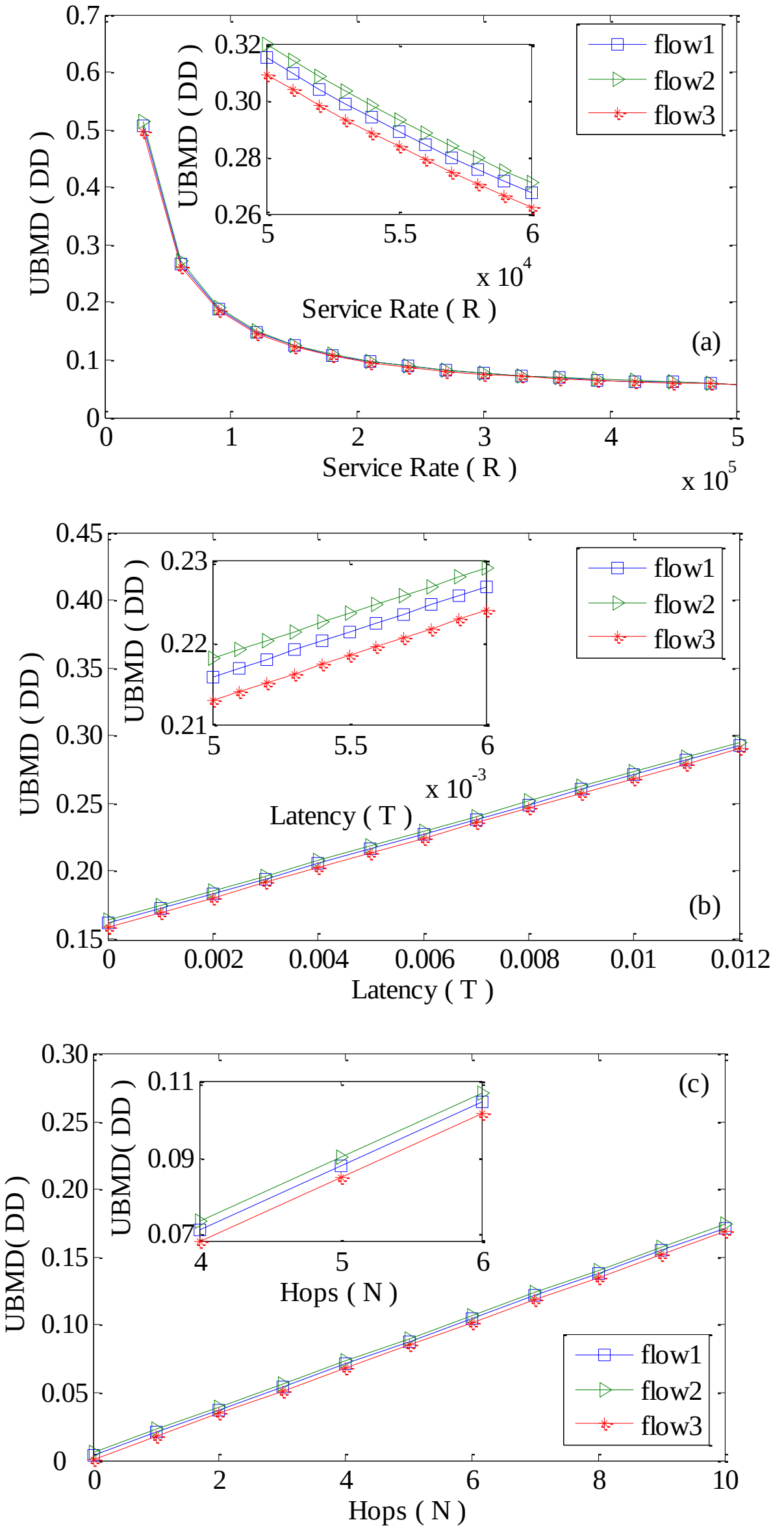

FIGURE 8: The upper bounds on multi-hops delay (UBMD) $DD$ in s: (a) $DD$ as a function of the service rate $R$ in Kbps for $T$=1, $d$=2 and $N$=10; (b) $DD$ as a function of the latency $T$ in s for $R$=100, $d$=2 and $N$=10; (c) $DD$ as a function of the hops $N$ for $R$=100, $T$=1 and $d$=2.

Figure 8(b) plots a *DD* curves as a function of the latency *T*. The upper bounds on *DD* increase in linear with the increasing of *T* values regardless of *R* values, *N* values and *d* values for certain flow. All the increase rates of *DD* are 11.

Figure 8(c) plots a *DD* curves as a function of the hops *N*. The upper bounds on *DD* increase in linear with the increase of *N* regardless of *R* values, *T* values and *d* values for certain flow. All the increase rates of *DD* are 0.017.

*Remark 8*: From Eqs. (17) and (19), we have the relation between the multi-hops delay jitter $\Delta D$ and the multi-hops delay *DD* as follows: $\Delta D = DD - \Sigma d$, where *d* is the fixed delay between two neighbor sensor nodes. As a result, we can get some numerical results about the upper bounds on $\Delta D$ by setting $d_1 = d_2 = \ldots = d_{N-1} = d = 0$, and the import of the service rate *R*, the latency *T* and the hops *N* on $\Delta D$ is similar to those on *DD*.

Figure 9 shows the import of the service rate *R*, the latency *T* and the hops *N* on the upper bounds on multi-hops effective bandwidth *ee*. We see a straightforward dependency: the upper bound on *ee* is larger for larger service rate; it is larger for smaller latency and smaller hops.

Figure 9(a) plots an *ee* curves as a function of the service rate *R*. The upper bounds on *ee* increase with the increase of *R* values, and the increase rate is getting smaller and smaller with the increase of *R* for certain flow regardless of the values of the latency *T*, the fixed delay *d* and the hops *N*. The import of the burst tolerance *b* on *ee* is more than that of the service rate *R* on *ee* for the high-values $R>30$ or the import of the service rate *R* is more. For example, if $N=10$, $T=1$, $R=20$ and $d=2$, the *ee* value of the flows: $A_1(t)$, $A_2(t)$ and $A_3(t)$ is 1.32, 1.26 and 0.30, and if $N=10$, $T=1$, $R=200$ and $d=2$, the *ee* value of the three flows is 4.98, 6.89 and 2.02, respectively.

Figure 9(b) plots an *ee* curves as a function of the latency *T*. The upper bounds on *ee* decrease with the increase of *T* values, and the decay rate is getting smaller and smaller with the increase of *T* for certain flow regardless of the values of the service rate *R*, the fixed delay *d* and the hops *N*. The changing tendency of *ee* for each flow is similar to that of *e*.

Figure 9(c) plots an *ee* curves as a function of the hops *N*. The upper bounds on *ee* decrease with the increase of *N* values. The decay rates of *ee* by the near exponential increase with the increase of the hops *N* for all flows, and they are smaller for larger burst tolerance *b* of the flows. For instance, if $N=1$, $R=100$, $T=1$ and $d=2$, the *ee* value of the flows: $A_1(t)$, $A_2(t)$ and $A_3(t)$ is 23.17, 30.48 and 11.21, and if $N=5$, $R=100$, $T=1$ and $d=2$, the *ee* value of the three flows is 56.19, 77.63 and 23.52, respectively.

In the case 1, assuming $N=2$, we can get the single-hop QoS. The study result shows the service rate *R* and the latency *T* produce the same effect on the upper bounds on single-hop delay *DD* and the multi-hops delay *DD*, and the single-hop effective bandwidth *ee* and the multi-hops effective bandwidth *ee*. If $R=100$ and $T=1$ and $d=2$, the upper bounds on single-hop delay *DD* of the flows: $A_1(t)$, $A_2(t)$ and $A_3(t)$ are 0.021, 0.023 and 0.018, and the upper bounds on single-hop effective bandwidth *ee* are 23.2, 30.5, and 11.2, respectively.

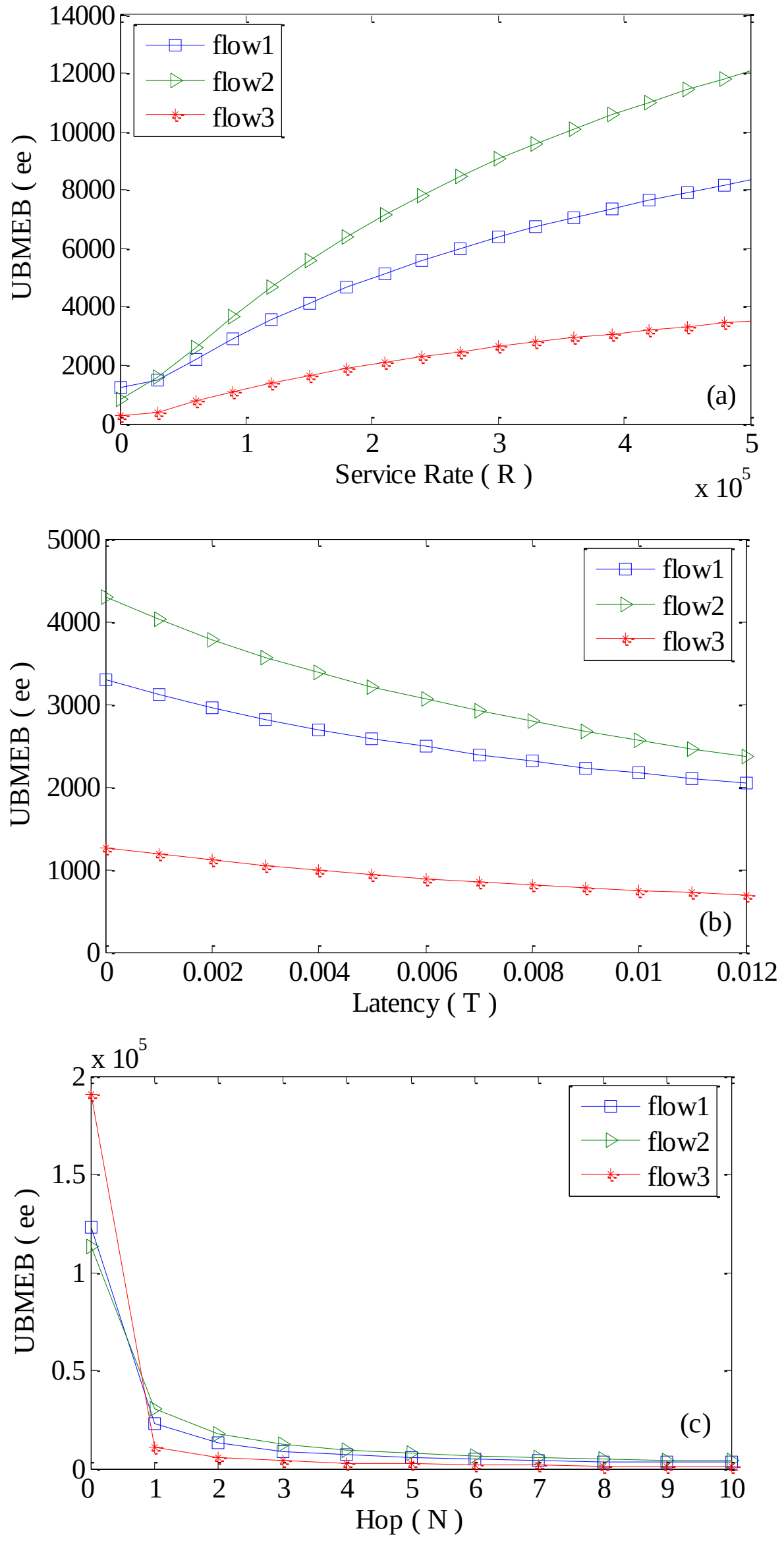

FIGURE 9: The upper bounds on multi-hops effective bandwidth (UBMEB) *ee* in Kb: (a) *ee* as a function of the service rate *R* in Kbps for $T=1$, $d=2$ and $N=10$; (b) *ee* as a function of the latency *T* in s for $R=100$, $d=2$ and $N=10$; (c) *ee* as a function of the hop *N* for $R=100$, $T=1$ and $d=2$.

To summarize, the performance curves denote the upper bounds of the single-hop/multi-hops QoS. In Figures 8 and 9, the curves show the deterministic worst-case end-to-end delay/effective bandwidth. It means that the values of the end-to-end delay must are lower than the values of the performance curves. We can reduce, by regulating the average arrival rate *r* and the burst tolerance *b* or the service rate *R* and the latency *T* of all sensor nodes on an end-to-end path, the values of the upper bounds on end-to-end delay to achieve this purpose that the end-to-end delay/jitter is very small. On the other side, we can increase, by regulating the average arrival rate *r* and the burst tolerance *b* or the service rate *R* and the latency *T*, the value of the upper bound on end-to-end effective bandwidth to gain a guaranteed bandwidth for those flows through the end-to-end path, and eventually reduce the end-to-end delay/jitter.

### 5.2.2. The case 2

The sensor nodes (node 1, node 2, …, node $N-1$, node $N$), given in Figure 4, have the different service curves: $\beta_1(t)\neq\beta_2(t)\neq\ldots\neq\beta_{N-1}(t)\neq\beta_N(t)$. By the number of the flows and the values of the average arrival rate and the burst tolerance of the arrival curves in Table 3, without any loss of generality, we assume the parameters of the service curves of the five sensor nodes (node 1, node 2, node 3, node 4, node 5), used for numerical calculation in the following, are given in Table 4.

TABLE 4: The parameters of the service curves

| Service curve | *R* (Mbps) | *T* (ms) |
|---|---|---|
| $\beta_1(t)$ | 540 | 5.80 |
| $\beta_2(t)$ | 510 | 7.80 |
| $\beta_3(t)$ | 624 | 3.38 |
| $\beta_4(t)$ | 480 | 6.54 |
| $\beta_5(t)$ | 420 | 3.20 |

The next, we calculate the upper bounds on multi-hops delay *DD*, the multi-hops delay jitter $\Delta D$ and the multi-hops effective bandwidth *ee* of the three flows (in Table 3) from the sensor node 1 to the sensor node 5. If the fixed delay between two adjacent sensor nodes $d_1$=1.2, $d_2$=2.3, $d_3$=2.0, $d_4$=3.5, $d_5$=2.6, from Eqs. (17), (19) and (20), we hold the *DD* value of the flows: $A_1(t)$, $A_2(t)$ and $A_3(t)$ is 58.9, 59.4 and 58.2, and the $\Delta D$ value of the three flows is 47.3, 47.8 and 46.6, and the *ee* value of the three flows is 8.15, 11.78 and 3.43, respectively.

This case 2 is the general form of the case 1. Using the same method, we calculate the upper bound on single-hop delay/jitter/effective bandwidth. The single-hop QoS is compared to the multi-hops QoS, and shows similar trend between them. But unlike the case 1, we should consider how to regulate the various service rate *R* and the various latency *T* of each sensor node on an end-to-end path. The aim is to reduce the values of the upper bounds on end-to-end delay/effective bandwidth, and get the tolerable delay/jitter for the tracking and environment scouting applications in the WSNs.

### 5.2.3. The case 3

To display the guaranteed QoS model appears to, presented in sections 3 and 4, be valid for the WSNs with self-similar traffic flows.

The method is as follows: we replace the simple leaky bucket regulators with the fractal leaky bucket regulators in the two-layer scheduling model of sensor nodes; the envelope of the fractal leaky bucket regulators are also expressed as Eq. (12) in Ref. [26], and the average arrival rate $r_{i,k}$ and the burst tolerance $b_{i,k}$ are interpreted as follows, respectively,

$$r_{i,k}=m_{i,k}+\sigma_{i,k}(1-H_{i,k})\sqrt{2\gamma\left(\frac{H_{i,k}}{1-H_{i,k}}\right)^{H_{i,k}-1}},$$

$$b_{i,k}=\sigma_{i,k}(1-H_{i,k})\sqrt{2\gamma\left(\frac{H_{i,k}}{1-H_{i,k}}\right)^{H_{i,k}}}.$$

Where $m_{i,k}$ is interpreted as the long-term average arrival rate of the self-similar traffic, $\sigma_{i,k}$ as the standard deviation, $H_{i,k}$ as the Hurst parameter with the values ranging from 0.5 to 1, and $\gamma$ is a positive constant of 6.

*Remark 9*: The fractal leaky bucket regulators and aggregators don't increase the upper bounds on buffer queue length/delay and the buffer requirements of a sensor node, and don't increase the upper bounds on single-hop/multi-hops delay/jitter/effective bandwidth of the WSNs.

To provide comparisons between performances with self-similar traffic flows and that with general traffic flows for the greater details in the research analysis, we consider the general scenario of the WSNs in Figure 4 and three self-similar traffic flows including six self-similar micro-flows. Without loss of generality, for any *i* and *k*, we assume the $m_{i,k}$ and $\sigma_{i,k}$ values of the self-similar micro-flows are equal to the $r_{i,k}$ and $b_{i,k}$ values of the micro-flows in Table 3, respectively.

Figure 10 shows the import of the Hurst parameter *H* on the upper bounds on multi-hops delay *DD* and multi-hops effective bandwidth *ee* in case of the self-similar micro-flows with the same Hurst parameters. We see a straightforward dependency: the upper bound on *DD* and *ee* are smaller for larger Hurst parameter.

The *DD* values and the *ee* values all decrease with the increasing of the *H* values under increasing rate. The import of the Hurst parameter on the *DD* values and the *ee* values increase with the increase of the standard deviation $\sigma_{i,k}$. For instance, if $H$=0.75, $R$=100, $T$=1, $d$=2 and $N$=10, the $DD$ and $ee$ value of the flows: $A_1(t)$, $A_2(t)$ and $A_3(t)$ is 215.9, 218.9 and 212.1, and 3.95, 5.26 and 1.55, and if $H$=0.95 and the same values of other parameters mentioned above, the $DD$ value and the $ee$ value of the three flows is 129, 131 and 127, and 2.34, 2.82 and 0.83, respectively.

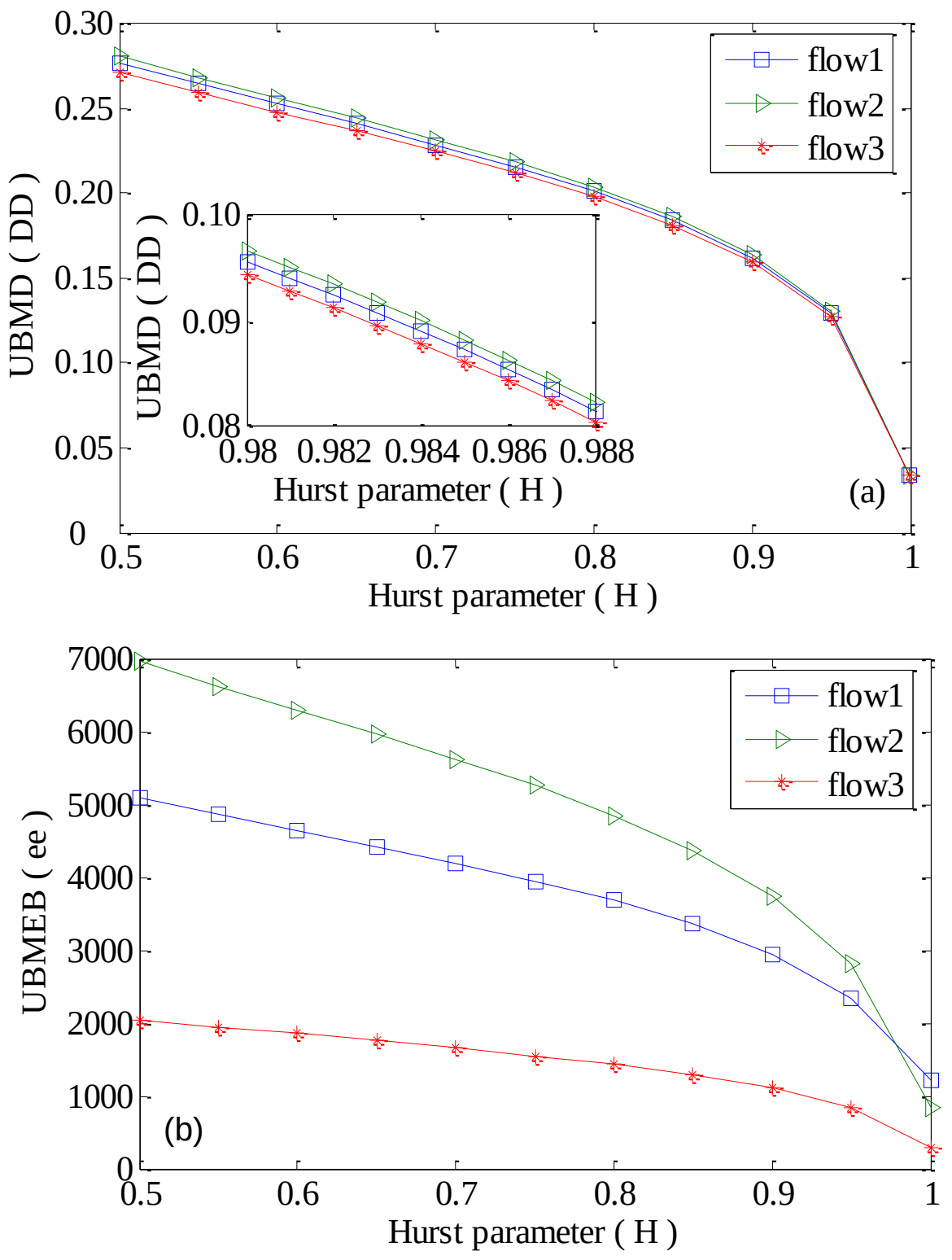

FIGURE 10: The upper bounds on multi-hops delay (UBMD) $DD$ in s and multi-hops effective bandwidth (UBMEB) $ee$ in Kb: (a) $DD$ as a function of the Hurst parameter $H$ for $R$=100, $T$=1, $d$=2 and $N$=10; (b) $ee$ as a function of the Hurst parameter $H$ for $R$=100, $T$=1, $d$=2 and $N$=10.

Now, we assume the Hurst parameters of the self-similar micro-flows have different values. For example, if $H_{1,1}$=0.90, $H_{1,2}$=0.80, $H_{1,3}$=0.75, $H_{2,1}$=0.85, $H_{2,2}$=0.60 and $H_{3,1}$=0.70, and $R$=100, $T$=1, $d$=2 and $N$=10, the $DD$ value and the $ee$ value of the flows: $A_1(t)$, $A_2(t)$ and $A_3(t)$ is 222, 226 and 218, and 3.76, 5.74 and 1.65, respectively.

Besides, calculation and analysis of the node QoS and the single-hop QoS, such as the upper bounds on buffer queue length/delay/effective bandwidth of a sensor node, and the single-hop delay/jitter/effective bandwidth, and the multi-hops delay jitter in the WSNs with self-similar traffic flows are done, and the results are similar.

In this special case with self-similar traffic flows, we can get the guaranteed QoS by regulating the service rate $R$ and the latency $T$ of sensor nodes, as can we do in the case 2. The difference is that we use the fractal leaky bucket regulators to regulate the arrival self-similar traffic flows. Obviously getting the Hurst parameter of the arrival self-similar traffic flow is a key in advance in the case. Then we can get the guaranteed QoS by regulating the average arrival rate $r$ and the burst tolerance $b$ of self-similar traffic flows in the WSNs.

*Remark 10*: Recently many related works on deterministic end-to-end delays have done. Lenzini and Mingozzi et al. computed the end-to-end delay based on the LUDB methodology [7], but the delay is the minimum among all the delay bounds and the LUDB methodology cannot be applied directly to non-nested tandems. Schmitt et al. achieved the worst-case end-to-end delays under blind multiplexing in tandem networks [8], and they dealt with arrival and service curves by a decomposition and recomposition scheme. Unlike Schmitt's method, Bouillard et al. directly computed the worst-case end-to-end delay instead of looking first for an end-to-end service curve by a decomposition and recomposition scheme [9], and may get a tighter bounds and a cheap complexity. Koubaa et al. proposed closed-form recurrent expressions for computing the worst-case end-to-end delays across any source-destination path in a cluster-tree WSN [17, 18, 19]. In this paper, the proposed network-calculus-based models is simpler for computing the upper bounds on buffer queue length/delay/effective bandwidth and multi-hops/single-hop delay/jitter/effective bandwidth using virtual buffer sharing in WSNs, and these models are suitable for various flows, including self-similar traffic flows.

In summary, based on the numerical results and analysis, we have found the parameters of the flow regulators and the service curves in the sensor nodes play an important role in modelling on a guaranteed QoS model for the WSNs, and got the following findings: (1) the upper bound on buffer queue length is smaller for larger service rate with high-values, and it is smaller for larger evolution time or larger Hurst parameter, and it is smaller for smaller latency or smaller service rate with low-values; (2) the upper bound on (multi-hops/single-hop) delay/jitter is smaller for larger service rate or larger Hurst parameter, and it is smaller for smaller latency or smaller hops; (3) the upper bound on (multi-hops/single-hop) effective bandwidth is larger for larger service rate, and it is smaller for larger latency or larger hops or larger Hurst parameter. In order to get network performance optimization and the guaranteed QoS of the WSNs, such as low delay for tracking, we can reduce the upper bounds on (end-to-end) delay/jitter or increase the upper bounds on (end-to-end) effective bandwidth by designing the rational regulator parameters, including the average arrival rate and the burst tolerance, and the rational scheduler parameters such as the service rate and the latency, of sensor nodes.

## 6. CONCLUSION

In this paper, we have discussed the problem of the guaranteed QoS for flows. Firstly, based the arrival curve and the service curve in the network calculus, we have presented the system skeleton, involving the sensor node model on virtual buffer sharing, the flow source model and the two-layer scheduling model of sensor nodes, and so on. Secondly, with the system skeleton, we have not only drawn the node QoS model, such as the upper bounds on buffer queue length/delay/effective bandwidth, but also drawn the single-hop/multi-hops QoS model, such as the upper bounds on single-hop/multi-hops delay/jitter/ effective bandwidth. Finally, we have shown the practicability and simplicity of the model and our approach using example results in the paper. We can optimize network performances by designing reasonable regulators and schedulers of the WSNs nodes. A network calculus approach is as a trade-off between complexity and accuracy. It is general, simple and practicable for provisioning the guaranteed QoS in the WSNs and other wireless networks with some characteristics of the distribution and the multi-hops.

Ongoing and future works include: (1) implementing the algorithmic build upon the proposed network-calculus-based model to ensure polynomial time complexity, for example, the computational complexity of node QoS algorithmic is $O(cn)$, where $c$ is the number of micro-flows, n is the number of flows, and the computational complexity of buffer queue length algorithmic is $O(c^2n^2)$, and the computational complexity of multi-hops/single-hop QoS algorithmic is $O(cnN)$, where $N$ is the number of hops; (2) investigating statistical sensor network calculus in order to capture the stochastic and dynamic behaviors of the WSNs.

## ACKNOWLEDGMENTS

This research was supported in part by a grant from the National Natural Science Foundation of China (60973129, 60903058 and 60903168), the China Postdoctoral Science Foundation funded project (200902324), the Specialized Research Fund for the Doctoral Program of Higher Education (200805331109), the Scientific Research Fund of Hunan Provincial Education Department of China (10B062), the Program for Excellent Talents in Hunan Normal University (ET10902) and the Startup Project for Doctoral Research Supported by Scientific Research Fund of Hunan Normal University (110608). The material in this paper was presented in part at the IEEE First International Conference on Communications and Networking in China Vehicular Technology Conference (ChinaCOM), Beijing, China, October 25-27, 2006, the IEEE International Workshop of Information Technology and Security (WITS), Shanghai, China, December 20-22, 2008, and the ISECS International Colloquium on Computing, Communication, Control, and Management (CCCM), Guangzhou, China, August 04-05, 2008. The authors would like to thank the reviewers for their valuable comments.